\documentclass[12pt,letterpaper]{extarticle}

\usepackage{harvard}
\usepackage{enumerate}
\usepackage{url}
\usepackage{framed}
\usepackage{float}
\usepackage{comment}
\usepackage{multirow}
\usepackage{graphicx}
\usepackage{caption}
\usepackage{subcaption}

\usepackage{marginnote}
\usepackage{setspace}

\floatstyle{ruled}
\newfloat{algorithm}{hpt}{lop}
\floatname{algorithm}{Algorithm}

\usepackage{amsthm}

\usepackage{amssymb}

\usepackage{booktabs}
\usepackage[table]{xcolor}

\makeatletter
\newcommand{\raisemath}[1]{\mathpalette{\raisem@th{#1}}}
\newcommand{\raisem@th}[3]{\raisebox{#1}{$#2#3$}}
\makeatother

\newif\ifhide

\usepackage{empheq}

\usepackage{pgffor}

\foreach \x in {A,...,Z}{%
  \expandafter\xdef\csname cal\x\endcsname{\noexpand\ensuremath{\noexpand\mathcal{\x}}}
}
\foreach \x in {A,...,Z}{%
  \expandafter\xdef\csname scr\x\endcsname{\noexpand\ensuremath{\noexpand\mathscr{\x}}}
}
\foreach \x in {A,...,Z}{%
  \expandafter\xdef\csname rm\x\endcsname{\noexpand\ensuremath{\noexpand\mathrm{\x}}}
}
\foreach \x in {A,...,Z}{%
  \expandafter\xdef\csname bb\x\endcsname{\noexpand\ensuremath{\noexpand\mathbb{\x}}}
}
\foreach \x in {A,...,Z}{%
  \expandafter\xdef\csname bf\x\endcsname{\noexpand\ensuremath{\noexpand\mathbf{\x}}}
}

\usepackage{color}

\definecolor{red}{rgb}{1,0,0}
\definecolor{gray}{rgb}{0.5,0.5,0.5}
\definecolor{darkgray}{rgb}{0.4,0.4,0.4}
\definecolor{blue}{rgb}{0,0,1}
\definecolor{green}{rgb}{0,1,0}

\makeatletter
\@ifclassloaded{report}{
   \newtheorem{theorem}{Theorem} }
{   }
\makeatother

\newtheoremstyle{mytheoremstyle}
{\topsep}                    
{\topsep}                    
{\itshape}                   
{}                           
{\bfseries \scshape}         
{.}                          
{.5em}                       
{}  

\theoremstyle{mytheoremstyle}

\usepackage{ifxetex}
\ifxetex
\usepackage{stix}
\else
\usepackage{lmodern}
\usepackage[T1]{fontenc}
\usepackage{mathrsfs}
\fi

\usepackage[letterpaper]{geometry}
\begin{document}

\title{\Large \bf  Do the Golden State Warriors Have Hot Hands?}

\author{Alon Daks
  ~ Nishant Desai  ~ Lisa R. Goldberg\footnote{All three authors are at University of California, Berkeley. Please contact {\tt lrg@berkeley.edu} for more information.}} 

\date{\today\footnote{The authors are grateful to  Bob Anderson,  Laurent El Ghaoui, Kellie Ottoboni, Ken Ribet, Stephanie Ribet and Paul Solli for contributions to this article.}}
\maketitle
\begin{abstract} 
 Star Golden State Warriors Steph Curry, Klay Thompson, and Kevin Durant are great shooters but they are not streak shooters. Only rarely do they show signs of a hot hand. 
 This conclusion  is based on an empirical analysis of field goal and free throw data from the 82 regular season and 17 postseason games played by the Warriors in  2016--2017.
 Our analysis is inspired by the iconic 1985 hot-hand study by Thomas Gilovitch, Robert Vallone and Amos Tversky, but uses a permutation test to automatically account for Josh Miller and Adam Sanjurjo's recent small sample correction. In this study we show how long standing problems can be reexamined using nonparametric statistics to avoid faulty hypothesis tests due to misspecified distributions.
 
\end{abstract}
\newpage

\begin{quote}

The true believer in the law of small numbers commits his multitude of sins against the logic of statistical inference in good faith.

\hfill  Amos Tversky and Daniel Kahneman, ``Belief in the Law of Small Numbers''
\end{quote}

\section{The original hot-hand study}

All basketball fans know about the hot hand: pass to a teammate on a scoring streak since her or his chances of making the next basket are higher than usual.

This venerated principle  was discredited  in 1985 by Amos Tversky and two co-authors, Thomas Gilovich and Robert Vallone.\footnote{Please refer to  \cite{hothand}.}  Their statistical  study of   field goal data from the Philadelphia 76ers, free throw data from the Boston Celtics, and a controlled 100-shot-per-player experiment on Cornell University varsity and junior varsity basketball players
seemed to prove that such scoring streaks are not out of the ordinary.   Although fans think that their players have hot hands, the streaks can be explained by mere chance.  

This game-changing news received a lukewarm reception from professional sports.  Red Auerbach, President of the Boston Celtics when the hot-hand study was released, famously gave his views on Tversky: ``Who is this guy? So, he makes a study.  I couldn't care less.''

Academics, however,  seemed to be fascinated by the finding. The 1985 study launched an avalanche of scholarly literature, and the hot hand question has propelled investigations about the conflict between the instincts of professionals and the cold hard facts of science. In his recent best-selling book, {\it The Undoing Project,}   Michael Lewis tells the story of  Amos Tversky and  his lifelong collaborator, Nobel Laureate Daniel Kahneman.  Their research prompted the study of behavioral economics  and it has transformed our understanding of the flaws in human decision making.  Lewis writes,  ``Tversky had the clear idea of how people misperceived randomness.  [...]  People had incredible ability to see meaning in these patterns where none existed.''

Early in their careers, Amos Tversky and Daniel Kahneman considered  the human tendency to draw conclusions based on a few observations, which they called ``the law of small numbers.''\footnote{Please refer to  \cite{small}.} This is a playful allusion to the law of large numbers, which provides guidance about when accurate inference can be drawn from a large data set.  There is no general rule about how to  draw inference from a small data set, and it can be difficult to notice that there is a problem. Even for an expert.

\section{A basic statistical formulation of a hot hand}

There are many ways to represent the notion of ``hot hand'' in a statistical experiment.   In this study as in the original study,   a {\it hot hand} is an abnormally high probability of making a shot, given a string of hits.  

This formulation abstracts away some of the details of the game. Some shots are harder to make than others, and defensive maneuvers by the opposition may  put  high performing player in a disadvantageous spot. Still, it is interesting to look at the results of this simple experiment before attempting to add realism.

There are also many ways that one might define the term ``abnormally high.''  Again we follow the original hot-hand study, which relied on a difference of conditional probabilities:  the probability of making a shot, given a string of hits minus the probability of making a shot, given an equally long string of misses.   If the observed difference is large relative to the typical difference corresponding to a random string of the same length and the same number of hits, the observation corresponds to a hot hand.

\section{A ``law of small numbers'' error in the original hot-hand study, and a correction}

In 2015, statisticians Josh Miller and Adam Sanjurjo documented an error in the original hot-hand study.\footnote{Please refer to \cite{hothand2}.}  Ironically, the error concerns the law of small numbers.

To understand the mistake, consider Klay Thompson's shooting record in the December 23, 2016 game against the Detroit Pistons. The record is represented by a string  of 1's (hits) and 0's (misses).
$$1110100110000011$$

Thompson took 16 shots in this game and as it happened, he made  exactly half of them.  We can look at our statistical formulation of the hot hand on this string. First, we compute the empirical probability of a hit, given two previous hits.   There are  four  instances of two hits in a row, indicated by the string ``11.''  We know what happened after the first  three instances --- Thompson hit the first time and missed the second and third times.  But nothing happened after the fourth  instance  of ``11'' because the game ended  before he could take another shot. We call the final  ``11'' an {\it unrealized conditioning set}, and it complicates the estimation of the conditional probabilities used in hot-hand studies.  
Perhaps the best we can say is that in the game under consideration,  we observed  Thompson scoring 1/3 of the time following two hits in a row.  In the other direction, given that Thompson missed twice in a row, he scored 2/5 of the time. The second calculation is more straightforward because there is no unrealized conditioning set.  The difference of the two conditional probabilities is  $1/3 - 2/5 = -1/15.$

Is this difference, $-1/15$, abnormally high?   Perhaps for this string,  which is half hits and half misses, there is a natural benchmark against which to measure ``abnormal'.'  Based on the data,  perhaps it is reasonable to assume that  the probability of making a shot after two hits is the same as the probability of making a shot after two misses:  50 percent.  Against this benchmark, the average difference in conditional probabilities is 0, which does not make $-1/15$ look abnormally high.  This benchmark is consistent with the original hot-hand study.

 However --- and this is the where the law of small numbers comes in --- the 50-50 benchmark is the wrong choice.  It would have been correct had we been dealing with infinite strings, but games don't go on indefinitely.  In practice, we deal with finite strings.  Many have unrealized conditioning sets, and some have no conditioning sets at all, so the  natural benchmark requires a small sample adjustment.   In a string of length 16 that is half 1's and half 0's, the probability of a hit following two hits is less than the probability of a hit following two misses:  reversals are more probable than continuations. That is the observation of Miller and Sanjurjo, and it is consistent with the gambler's fallacy, the impression that a reversal in fortunes is ``due.'' Taking this phenomenon into account, we realize that the expected difference in conditional probabilities is some value less than zero. This leaves open the potential for a study using the incorrect 50-50 null hypothesis to fail to reject the null in cases where a correctly specified null distribution would lead to a rejection in favor of the right-sided alternative.

The no-hot-hand conclusion in the original study was based on a statistically insignificant difference between the observed data and the erroneous benchmarks.   When the required adjustment was applied to the controlled 100-shot-per-player experiment on Cornell University players, Miller and Sanjurjo report that the no-hot-hand finding was reversed in several cases.

\section{In search of the Warriors' hot hand}
As noted by Miller and Sanjurjo, a permutation test of an observed string of hits and misses automatically implements the small sample correction.\footnote{Please see \cite[Section 3.1]{hothand2}.}  In this test, a property of a particular string of zeroes and ones is compared to the same property in random rearrangements of the entries of the string.  This allows for a quantitative assessment of a property's rarity.

Here, we  use the permutation test to decide in which games Curry, Thompson, and Durant had hot hands.  We also investigated the hot handedness of the Warriors, quarter by quarter.

\subsection{Data}

\begin{table}[h]
\centering
\begin{tabular}{ l  c  c  c  c}
&  Curry &  Thompson & Durant & Warriors \\
\toprule
Games & 96 & 95  & 77 & 99\\ \midrule
Observations &  96  &  95 & 77 & 396 \\ \midrule
Season Percentage  &56\% & 51\% & 63\%  & 56\%   \\ \midrule
Average Game Percentage & 56\% & 50\% & 62\%  & 56\%\\ \midrule
StDev Game Percentage & 11\% &12\%&12\% & 10\%\\ \midrule
Average Number of Shots & 24 & 20 & 23 & 28 \\ \midrule
StDev Number of Shots & 5&5&7&4 \\
 \bottomrule
\end{tabular}
\caption{Summary statistics for the 2016--2017 regular and post season games played by the Warriors.  The shot patterns we analyze include field goals and free throws. \label{summary}}
\end{table}

For Curry, Thompson, and Durant,  we compiled a string of 1's and 0's representing hits and misses for each of the  99  regular and postseason games that they played in 2016--2017.   Curry played 96 games, Thompson played 95, and Durant played 77.  We also compiled the strings of hits and misses for the Warrior team, quarter by quarter, leading to $396 = 99 \times 4$ quarters over the season.

\subsection{Experimental design and test statistics}

An observation $X$ is a game-long string of hits and misses for Curry, Thompson, and Durant or a quarter-long string of hits and misses for the Warriors.  The string includes both field goals and free throws.  We use a permutation test to determine whether the observation showed evidence of a hot hand.

For an observed string, we computed a test statistic, $t_k$,  the conditional fraction of hits, given $k$ prior hits  less the conditional fraction of hits given $k$ prior misses, where $k$ equals 1, 2 or 3.  Then we permuted the string of 0's and 1's representing the shot pattern 10,000 times and computed $t_k$ on each permutation.

Mathematically, the test statistic $t_k$ on a string $X$ of length $L$ is defined as follows:
\begin{align}
t_k(X) &=  \frac{1}{H_k}\sum_{\tau = k + 1}^{L} I \left(X_\tau =1  \biggr\rvert   \prod_{u = \tau - k }^{\tau -1}  X_u  = 1\right)- \frac{1}{M_k} \sum_{\tau = k + 1}^{L} I\left(X_\tau =1  \biggr\rvert   \prod_{u = \tau - k}^{\tau -1} \left(1 -  X_u\right)  = 1\right) \nonumber\\
      &=  \hat P \left(X_\tau =1  \biggr\rvert   \prod_{u = \tau - k}^{\tau -1}  X_u  = 1\right) - \hat P\left(X_\tau =1  \biggr\rvert   \prod_{u = \tau - k}^{\tau -1} \left(1 -  X_u\right)  = 1\right)\nonumber\\
      &= t_{k, hit}(X) - t_{k,miss}(X),  \label{hitmiss}
\end{align}
where  $H_k$ and $M_k$ are the numbers of substrings of $k$ hits and $k$ misses that are followed by shots, $X_\tau$ is the $\tau$th entry of $X$, $\hat P$ is the empirical probability.  The value $k$ is the depth of the conditioning set.

The fraction of permuted test statistics that exceed the observed test statistic is its $p$-value.\footnote{In practice, the value of the test statistic on many of the permuted strings is the same as the value of the test statistic on the observed shot pattern. Therefore, there is some latitude in how to to define the $p$-value. Our choice makes it as easy as possible to reject the null hypothesis of ``no hot hand.''}   A smaller $p$-value corresponds to stronger evidence of a  hot hand.

\subsection{Results}
Statistics describing the number of shots taken and hit frequency are in Table~\ref{summary}. 
The results shown below are mostly about $t_2$.  Results for $t_1$ and $t_3$ are qualitatively similar.

\begin{figure}[H]
\centering
\begin{subfigure}{0.5\textwidth}
  \centering
  \includegraphics[width=\linewidth]{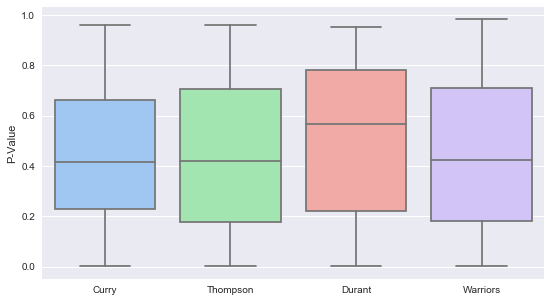}
  \caption{Box-plots}
  \label{fig:sub1}
\end{subfigure}%
\begin{subfigure}{0.5\textwidth}
  \centering
  \includegraphics[width=\linewidth]{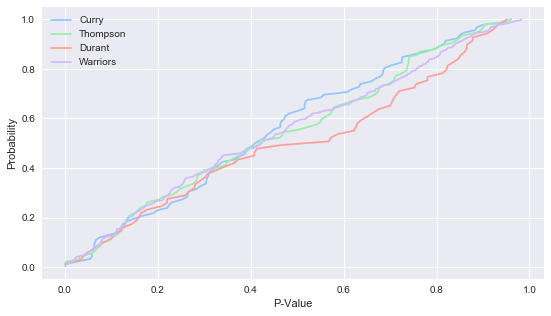}
  \caption{Cumulative distributions}
  \label{fig:sub2}
\end{subfigure}
\caption{$p$-values for permutation tests with  a depth-2 conditioning set:  the conditional number of hits given two prior hits minus the conditional number of hits given two prior misses.}
\label{boxcum2}
\end{figure}

Box plots and cumulative distributions of $p$-values for the test statistic $t_2$ calculated for the 2016--2017 season games played Curry, Thompson, and Durant and for the quarters played by the Warriors, are shown in Figure~\ref{boxcum2}.  Few observations are significant at the 5\% level.

Table~\ref{output} displays the number of observations that are significant at the 5\% level for conditioning sets of depths 1, 2 and 3.

\begin{table}[h]
\centering
\begin{tabular}{ l  c  c  c  c}
&  Curry &  Thompson  & Durant & Warriors \\
\toprule
Games & 96 & 95  & 77 & 82\\ \midrule
Observations &  96  &  95 & 77 & 396 \\ \midrule
Conditioning set  \\
depth\\
\hspace{3mm} 1 & 7 & 5 & 3 & 26 \\
\hspace{3mm} 2 & 2 & 4 & 3 & 21 \\
\hspace{3mm} 3 & 3 & 2  & 3 & 30\\
 \bottomrule
\end{tabular}
\caption{Number of hot hand observations at the 5\% significance level with a depth-2 conditioning set. \label{output}}
\end{table}

In addition to the $t_k$ statistic from the original hot-hand study, we considered a number of other statistics that could potentially be indicative of a hot hand. We looked at only the left summand of $t_k$, $t_{k,hit}$. In this case, we tested a simpler definition of the hot hand: a player has a hot hand if she or he has a higher chance of making a shot following a string of successful shots. The notion of ``higher chance'' here is defined the same way as in the above test.

For both of our test statistics, we also considered two additional nonparametric tests. First, instead of permuting a shot string for a given game, we estimated the distribution of the test statistic by sampling 10,000 new binary strings by simulating $n$ Bernoulli($p$) trials, where $p$ is the player's shooting percentage for the game in question and $n$ is the number of shots the player took. In the second test, we again simulated $n$ Bernoulli trials, however we let $p$ equal the player's season shooting percentage up to the start of the game being tested. Both alternative formulations yielded the same conclusion: no evidence of a hot hand.

\subsection{An example}

On December 5, 2016, Thompson  scored 60 points against the Los Angeles Clippers.  
He made 31 of the 44 shots he took, and his record for the game is shown below.
$$X(60) = 11011110010111111001110111101110111101010101$$
Does this indicate a hot hand?   
There are  19 instances of the string ``11,'' and they are followed by hits in 12 of the 19 cases.  There are two instances of the string ``00,'' and they are both followed by hits.
So  $t_2 = 12/19 - 1 = -7/19.$

\begin{figure}[H]
\centering
\includegraphics[scale=.75]{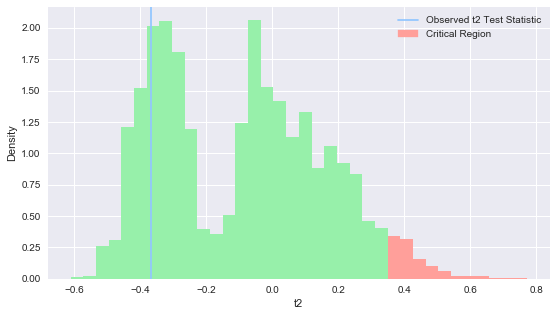}
\caption{$t_2$ statistics for Thompson's 60-point game against the Los Angeles Clippers on December 5, 2016. The green region represents the  null hypothesis, the histogram of values of $t_2$ corresponding to 10,000 permutations of the observed string.   The blue line marks the value of $t_2$ for the observed string, and the red critical region corresponds to the highest 5\% of values of $t_2$ --- the hot hand.}
\label{t2histogram} 
\end{figure}

Figure~\ref{t2histogram} shows the histogram of $t_2$ statistics for 10,000 permutations of  Thompson's $X(60)$ string.  The green region represents the  null hypothesis, the histogram of values of $t_2$ corresponding to 10,000 permutations of the observed string. In Appendix~\ref{null},  we explore the bimodality of the null hypothesis as well as other irregularities in its shape.   The blue line marks the value of $t_2$ for the observed string, and the red critical region corresponds to the highest 5\% of values of $t_2$ --- the hot hand.

This observation is exceptional for its length, and it is exceptional for its percentage of hits, $31/44  \approx 70\%$.  But the difference of conditional probabilities, $-7/19$, had a $p$-value of $0.84$, which is not exceptional at all. 

\section{Summary}

We examined the 2016--2017 regular season shooting records of Splash Brothers Steph Curry and Klay Thompson as well as the 2017 Finals MVP Kevin Durant. It is a magical experience to watch  these players on the court. When they are on a roll, they seem to be the essence of hot handedness. But our statistical study tells a different story. It indicates that in most of the 2016--2017  regular season games, they were not streak shooters --- they did not have hot hands.  So our conclusion, after adjusting for the small sample effect, is similar to the original conclusion, which did not account for the small sample effect. 

Of course,  this is not the end of the story.  Hot hands have long fascinated sports professionals and researchers, and we are not close to consensus on the right way to think about the issue. However, every empirical hot-hand study will rely on a finite data set, so small sample effects are bound to play a role in any correct interpretation of the results.  

Amos Tversky was 59 years old when he died in 1996, five years before the Nobel Prize that he would surely have shared with Daniel Kahneman was awarded, and almost two decades before the error in his hot-hand study was found. An awesome researcher and a huge basketball fan, he would no doubt be pleased about the correction of his error if he were with us today, and he would surely be watching the spellbinding Golden State Warriors.

\appendix
\section{Deconstructing the null hypothesis for $t_2$}\label{null}

\begin{figure}[H]
\centering
\includegraphics[scale=.75]{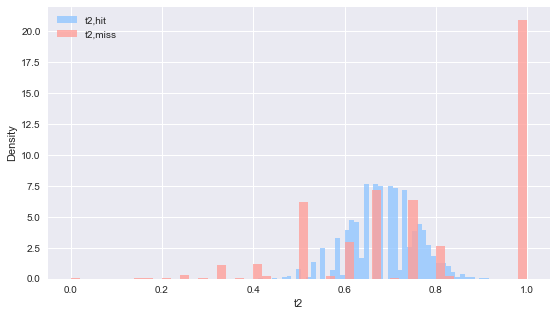}
\caption{$t_{2, hit}$  and   $t_{2, miss}$ histograms for Thompson's 60-point game against the Los Angeles Clippers on December 5, 2016.}
\label{t2histdec} 
\end{figure}

Here we look more closely at the null hypothesis for Thompson's 60-point game against the LA clippers, depicted in Figure~\ref{t2histogram}.
Formula~(\ref{hitmiss}) expresses  $t_2$ as a  difference of probabilities:
$$t_k =  t_{k, hit} - t_{k,miss}.$$
In Figure~\ref{t2histdec}, we show histograms for $t_{k, hit}$ and $t_{k,miss}$.  The shapes of these histograms  are determined entirely by the length of the string, 44, and the percentage of hits, $31/44\approx 70\%$. The spike at 1 for $t_{2, miss}$ occurs because missed shots are relatively rare:  in many strings of length 44 with 31 hits, all consecutive pairs of misses that are followed by anything at all are followed by a hit.

For longer strings and hit probabilities in the neighborhood of 50\%, the distribution of $t_2$ and its components tends to be unimodal and symmetric. But for relatively small strings corresponding, for example, to the number of shots a top professional basketball takes in a single game, histograms representing the null hypotheses for $t_2$ and its components are irregular. Consequently, finite sample methods of the type used in this article may be preferred to asymptotic results.

\bibliographystyle{jmr}
\bibliography{biblio}

\end{document}